\documentclass[%
longbibliography,
 reprint,
 showkeys,
superscriptaddress,
 amsmath,amssymb,
 pre,
]{revtex4-1}

\usepackage{graphicx}
\usepackage{dcolumn}
\usepackage{bm}
\usepackage[utf8]{inputenc}
\usepackage{graphicx}
\usepackage{color}
\usepackage{amsmath}
\usepackage{amssymb}
\usepackage{subfigure}
\usepackage{epsfig}
\usepackage{float}

\usepackage{lipsum}

\newcommand{\D}{\boldsymbol{D}}

\newcommand{\Gama}{\boldsymbol{\Gamma}}
\newcommand{\nd}{\boldsymbol{S}}

\newcommand{\M}{\boldsymbol{M}}

\newcommand{\B}{\boldsymbol{B}}

\newcommand{\nn}{\boldsymbol{n}}
\newcommand{\J}{\boldsymbol{J}}

\newcommand{\vv}{\boldsymbol{v}}
\newcommand{\rr}{\boldsymbol{r}}

\newcommand{\xxi}{\boldsymbol{\xi}}

\newcommand{\R}{\boldsymbol{R}}

\begin{document}



\title{Nondiffusive Fluxes in Brownian System with Lorentz Force}

\author{I.~Abdoli}
\affiliation{Leibniz-Institut  f\"ur Polymerforschung Dresden, Institut Theorie der Polymere, 01069 Dresden, Deutschland}

\author{H.D.~Vuijk}
\affiliation{Leibniz-Institut  f\"ur Polymerforschung Dresden, Institut Theorie der Polymere, 01069 Dresden, Deutschland}

\author{J.U.~Sommer}
\affiliation{Leibniz-Institut  f\"ur Polymerforschung Dresden, Institut Theorie der Polymere, 01069 Dresden, Deutschland} \affiliation{Technische Universit\"at Dresden, Institut f\"ur Theoretische Physik, 01069 Dresden, Deutschland}

\author{J.M.~Brader}
\affiliation{Department de Physique, Universit\'e de Fribourg, CH-1700 Fribourg, Suisse}

\author{A.~Sharma}
\email{sharma@ipfdd.de}
\affiliation{Leibniz-Institut  f\"ur Polymerforschung Dresden, Institut Theorie der Polymere, 01069 Dresden, Deutschland} \affiliation{Technische Universit\"at Dresden, Institut f\"ur Theoretische Physik, 01069 Dresden, Deutschland}

\begin{abstract}
The Fokker-Planck equation provides  complete statistical description of a particle undergoing random motion in a solvent. In the presence of Lorentz force due to an external magnetic field, the Fokker-Planck equation picks up a tensorial coefficient, which reflects the anisotropy of the particle's motion. 
This tensor, however, can not be interpreted as a diffusion tensor; there are antisymmetric terms which give rise to fluxes perpendicular to the density gradients. Here, we show that for an inhomogeneous magnetic field these nondiffusive fluxes have finite divergence and therefore affect the density evolution of the system. Only in the special cases of a uniform magnetic field or carefully chosen initial condition with the same symmetry as the magnetic field can these fluxes be ignored in the density evolution.  

\end{abstract}


\maketitle

\section{Introduction}
\label{section01}

While the effect of the Lorentz force on the properties of materials has been thoroughly studied in the context of
solid-state physics, much less is known about its influence on soft-matter systems which are dominated by overdamped dynamics. This becomes particularly interesting in light of the recent finding that the Lorentz force induces a particle flux perpendicular to density gradients, thus preventing a diffusive description of the dynamics~\cite{chun2018emergence,vuijk2019anomalous}. In this paper, we study the unusual fluxes induced by the Lorentz force and their effect on the nonequilibrium dynamics of the system.

Consider a single charged Brownian particle in a magnetic field $\B(\rr)$. Let $\nn$ be the unit vector in the direction of the magnetic field, and $B(\rr)$ be the magnitude. Due to the magnetic field, the particle is subjected to the Lorentz force $q\vv \times \B(\rr)$, where $\vv$ is its velocity. The dynamics of the particle are described by the following Langevin equation~\cite{langevin1908theory}:
\begin{align}
\dot \rr(t) &= \vv(t), \nonumber \\
m\dot \vv(t) &= -\gamma \vv + q \vv \times \B(\rr) + \sqrt{2\gamma k_B T}\xxi(t),
\label{LangevinB}
\end{align}
where $\rr$ is the position of the particle, $m$ is the mass of the particle, $q$ is the charge, $\gamma$ is the friction coefficient, $k_B$ is the Boltzmann constant, $T$ is the temperature and $\xxi(t)$ is Gaussian white noise with zero mean and time correlation $\langle\xxi(t)\xxi^{T}(t')\rangle=\boldsymbol{1}\delta(t-t')$. One can rewrite the Langevin equation in terms of the position dependent matrix $\Gama(\rr) = \gamma \boldsymbol{1} + qB(\rr)\M$ as
\begin{align}
m\dot \vv(t) &= -\Gama(\rr) \vv + \sqrt{2\gamma k_B T}\xxi(t), 
\label{LangevinB2}
\end{align}
where $\M$ is a matrix with elements $M_{\alpha \beta } = -\epsilon_{\alpha \beta \nu}n_{\nu}$, with $\epsilon_{\alpha \beta \nu}$ the totally antisymmetric Levi-Civita symbol in three dimensions and $n_{\nu}$ is $\nu$-component of $\nn$ for the Cartesian index $\nu$.

Often one is interested in the slow degree of freedom, which in this case is the position of the particle. The time scale at which velocity correlations decay, $\tau = m/\gamma$, is generally much smaller than the diffusion time scale of the particle. Since for times $t >> \tau$ the velocity decouples with the position, an equation for the position degrees of freedom alone can be obtained. This equation, referred to as the overdamped equation, is obtained by taking the small-mass limit of Eq.~\eqref{LangevinB}~\cite{gardiner2009stochastic,hottovy2015smoluchowski, volpe2016effective,vuijk2018pseudochemotaxis}. It has become common practise to start with the overdamped equation of motion as the model of the system under study~\cite{cates2013active, farage2015effective, sharma2016communication, stenhammar2016light, sharma2017escape, sharma2017brownian}. 
Whereas the Langevin equations are convenient for simulations, a statistical description is often preferred for theoretical analysis. To this end one derives the Fokker-Planck equation for the position degrees of freedom, which for a Brownian particle subject to inhomogeneous Lorentz force, is given as~\cite{hottovy2015smoluchowski, vuijk2019anomalous}

\begin{equation}
\label{FPE}
\frac{\partial P(\rr, t)}{\partial t} = \nabla \cdot \big[\D(\rr) \nabla P(\rr,t)\big],
\end{equation}
where $P(\rr,t)$ is the probability density and the tensor $\D(\rr)$ is given as
\begin{align}
\label{tensord}
\D(\rr) & =  D\left[\left(\boldsymbol{1} + \frac{\kappa^2(\rr)}{1 + \kappa^2(\rr)}\M^2\right) - \frac{\kappa(\rr)}{1+\kappa^2(\rr)} \M \right]  \\
        & = \D_s(\rr) + \D_a(\rr), \nonumber
\end{align}
where $D = k_B T/\gamma$ is the coefficient of a freely diffusing particle and $\kappa(\rr) = qB(\rr)/\gamma$ is a parameter quantifying the strength of Lorentz force relative to frictional force~\cite{vuijk2019lorenz}. $\D_s$ and $\D_a$ are the symmetric and antisymmetric parts of the tensor $\D$, which refer to the first and second terms of Eq.~\eqref{FPE}, respectively. The flux in the system is~\cite{chun2018emergence,vuijk2019anomalous}
\begin{equation}
\label{flux}
\J(\rr, t) = - \D(\rr)\nabla P(\rr, t). 
\end{equation}

The symmetric tensor $\D_s$ is the diffusion tensor and gives rise to the diffusive fluxes $\J_s(\rr, t) = - \D_s(\rr)\nabla P(\rr, t)$ along the density gradient. It is known that the motion of a Brownian particle is anisotropic in presence of magnetic field~\cite{balakrishnan2008elements}. The components of $\D_s$ along and perpendicular to the field are $D$ and $D/(1+\kappa^2)$, respectively.

The Fokker-Planck equation~\eqref{FPE} is unusual due to the presence of the antisymmetric tensor $\D_a$. This tensor captures the physical property of the Lorentz force that it curves the trajectory of a moving charge without performing work on it. More precisely, there exist fluxes $\J_a(\rr, t) = - \D_a(\rr)\nabla P(\rr, t)$, which are perpendicular to the density gradient~\cite{chun2018emergence,vuijk2019anomalous}. The presence of these nondiffusive fluxes makes the dynamics fundamentally different from purely diffusive. In this paper, we study the effect of these fluxes on the nonequilibrium dynamics of a system whose probability density evolves according to Eq.~\eqref{FPE}. We show that for inhomogeneous magnetic field, nondiffusive fluxes can significantly affect the density evolution. This is the main result of the paper. Only in the special cases of the uniform magnetic field or carefully chosen initial condition with the same symmetry as the magnetic field, can these fluxes be ignored in the dynamics of the probability distribution.

We have intentionally omitted writing overdamped equation obtained from the small-mass limit of Eq.~\eqref{LangevinB}. The overdamped equation has been derived in the past and shown to describe accurately the position statistics of the particle~\cite{hottovy2015smoluchowski, cerrai2011small, freidlin2012perturbations}. However, it has been shown recently that the overdamped equation is not suited for velocity-dependent quantities like flux and entropy production~\cite{vuijk2019anomalous}. This is a subtle consequence of the small-mass limit~\cite{hottovy2015smoluchowski} and was the main focus of Ref.~\cite{vuijk2019anomalous}. The Fokker-Planck equation, however, provides an accurate description of both the position statistics and the fluxes in the system. 

We take the following approach: We initialise the system into a nonequilibrium configuration and let it evolve in time according to Eq.~\eqref{FPE}. We numerically obtain the total fluxes, diffusive fluxes, and nondiffusive fluxes. We then only retain the symmetric part of the tensor $\D$ and let the system evolve in time. Since in the latter the nondiffusive fluxes are ignored, a comparison of the density evolution provides insight into the role of these fluxes in the dynamical evolution of the system.

\begin{figure*}
\centering
\resizebox*{15cm}{!}{\includegraphics{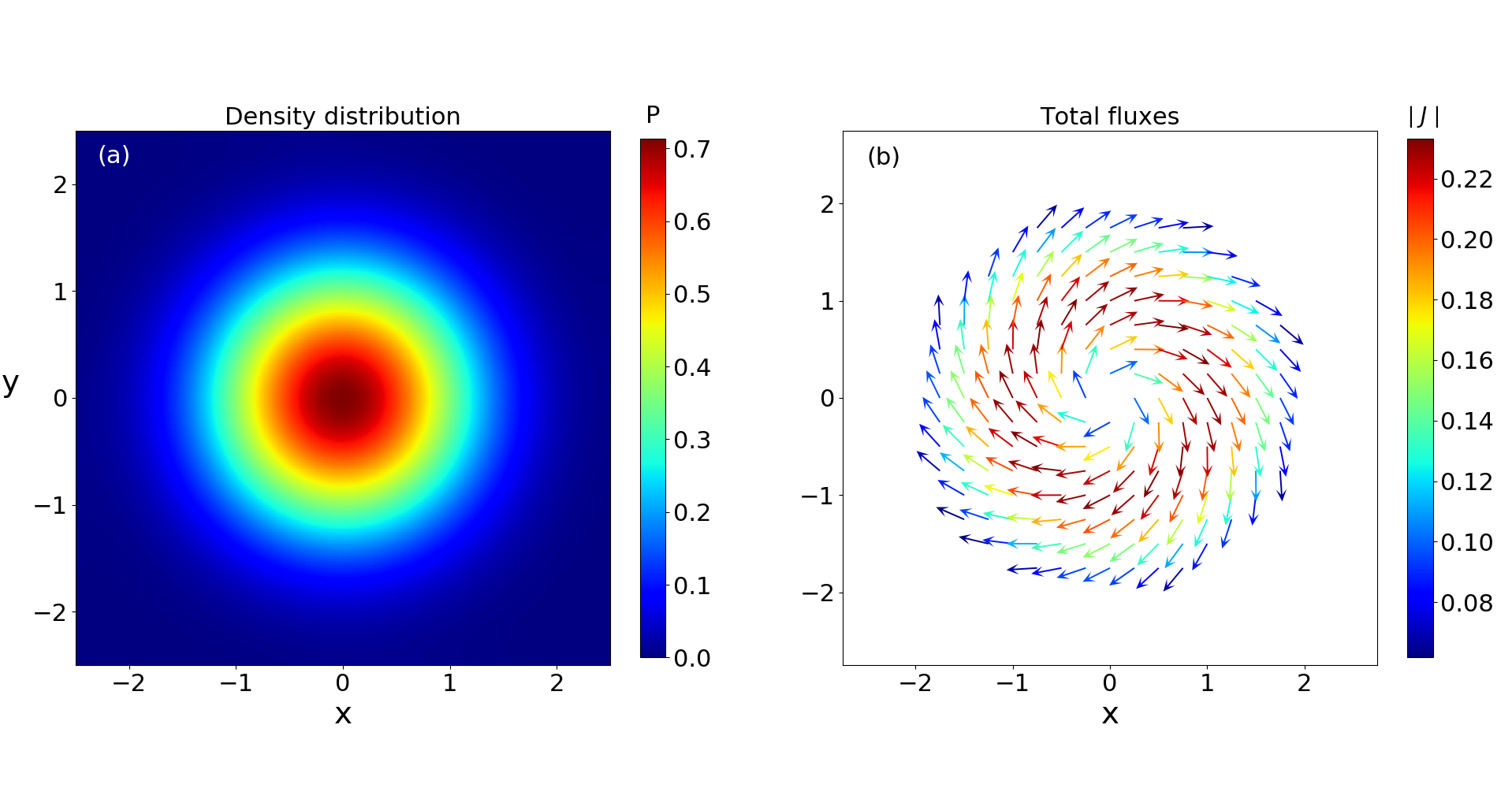}} \\ \vspace*{-6mm}
\resizebox*{15cm}{!}{\includegraphics{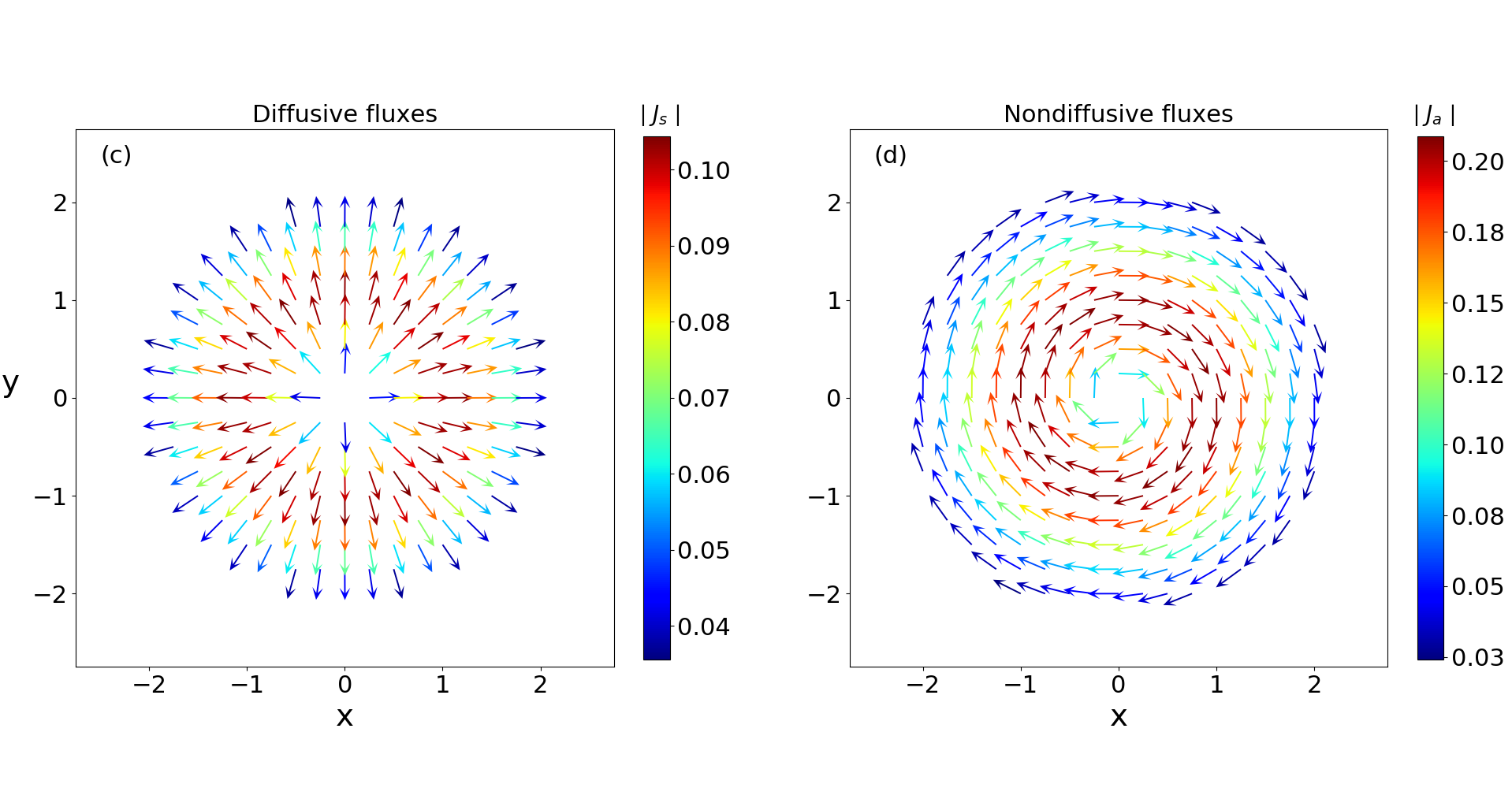}} \\ \vspace*{-0mm}
\caption{Constant magnetic field. (a) Density distribution in the system at time $t=1.0$ obtained from Eq.~\eqref{FPE} with $\kappa = 2$. The system size is $5 \times 5$. (b) Total fluxes in the system (Eq.~\eqref{flux}). The decomposition of the total flux into diffusive and nondiffusive components is shown in (c) and (d), respectively. 
The nondiffusive fluxes are perpendicular to the density gradients and purely rotational. Since these fluxes are divergence free, they do not affect the time evolution of the probability density.}
\label{constantB}
\end{figure*}
 

\section{Nondiffusive Fluxes in Magnetic Field}
\label{section02}

Here we consider a system with reflecting boundaries; The particles can not escape the confining geometry $\R$, hence we require $\nn_b\cdot\J(\rr, t)=0$ for $\rr\in\nd$, the boundary of $\R$, where $\nn_b$ is normal to $\nd$~\cite{gardiner2009stochastic}. The magnetic field is applied in the $z$ direction. Because the Lorentz force does not affect the motion in the $z$ direction, we restrict our analysis to the motion in the $xy$ plane. This effectively reduces the problem to two dimensions. All the results are obtained by numerically solving the Fokker-Planck equation~\eqref{FPE}. Using a central difference method we discretise the Fokker-Planck equation as follows:

\begin{align}
\label{FPExy}
&\frac{P(i\Delta x, j\Delta y; t+\Delta t) -P(i\Delta x, j\Delta y; t)}{\Delta t} = \\ \nonumber
& -\bigg[ \frac{J_x\big((i+1)\Delta x, j\Delta y; t\big) -J_x\big((i-1)\Delta x, j\Delta y; t\big)}{2\Delta x} \\ \nonumber
& + \frac{J_y\big(i\Delta x, (j+1)\Delta y; t\big) -J_y\big(i\Delta x, (j-1)\Delta y; t\big)}{2\Delta y}\bigg]
\end{align}
where $i$ and $j$ are integers, $\Delta x$ and $\Delta y$ are, respectively, the grid sizes in $x$ and $y$ directions, and $\Delta t$ is the integration time step which is fixed to $\Delta t = 5 \times 10^{-5}$ in this paper. At the boundaries of the system we use forward and backward differences. Our system is a rectangular box of size $L_x\times L_y$, where $L_x$ and $L_y$ are the lengths in $x$ and $y$ directions, respectively. Similarly, the fluxes are calculated numerically as 
\begin{subequations} \label{fluxxy}
\begin{align}
  &   J_x(i\Delta x, j\Delta y; t) = \\ \nonumber 
  & -\bigg[D_{xx} \frac{P\big((i+1)\Delta x, j\Delta y; t\big) - P\big((i-1)\Delta x, j\Delta y; t\big)}{2\Delta x} \\ \nonumber 
     & + D_{xy} \frac{P\big(i\Delta x, (j+1)\Delta y; t\big) -P\big(i\Delta x, (j-1)\Delta y; t\big)}{2\Delta y}\bigg]
     \label{fluxx}
\end{align}
\begin{align}
   &  J_y(i\Delta x, j\Delta y; t) = \\ \nonumber 
   &-\bigg[D_{yx}  \frac{P\big((i+1)\Delta x, j\Delta y; t\big) - P\big((i-1)\Delta x, j\Delta y; t\big)}{2\Delta x} \\ \nonumber 
     & + D_{yy} \frac{P\big(i\Delta x, (j+1)\Delta y; t\big) -P\big(i\Delta x, (j-1)\Delta y; t\big)}{2\Delta y}\bigg]
     \label{fluxy}
\end{align}
\end{subequations}

Here $D_{xx}$ and $D_{yy}$ are the symmetric diagonal terms which together with the antisymmetric terms $D_{xy}$ and $D_{yx}$ form tensor $\D$. 
For the numerical calculations we fix the diffusion coefficient $D = 1$ and $\Delta x = \Delta y = 0.01$  throughout this paper. Time is measured in units of  $\gamma/k_BT$, which is the time the particle takes to diffuse over a unit distance.

After initialising the system into a nonequilibrium configuration, we let it evolve in time according to Eq.~\eqref{FPE} and numerically obtain the fluxes (i.e., the total fluxes, diffusive and nondiffusive fluxes) in the system. 



\subsection{Constant Magnetic Field}
\label{subsectionA} 

\begin{figure*}
\centering
\resizebox*{15cm}{!}{\includegraphics{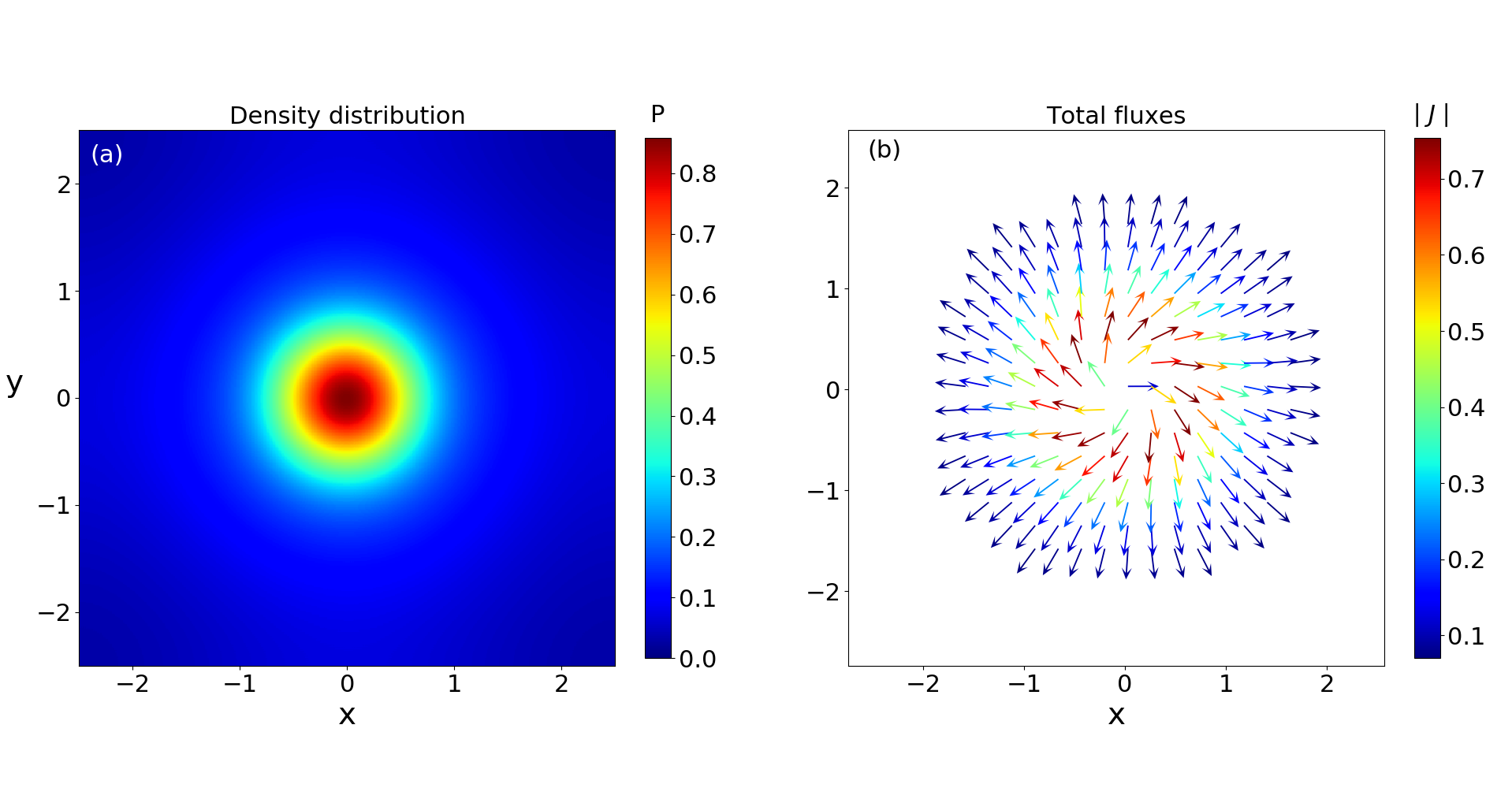}} \\ \vspace*{-2mm}
\resizebox*{15cm}{!}{\includegraphics{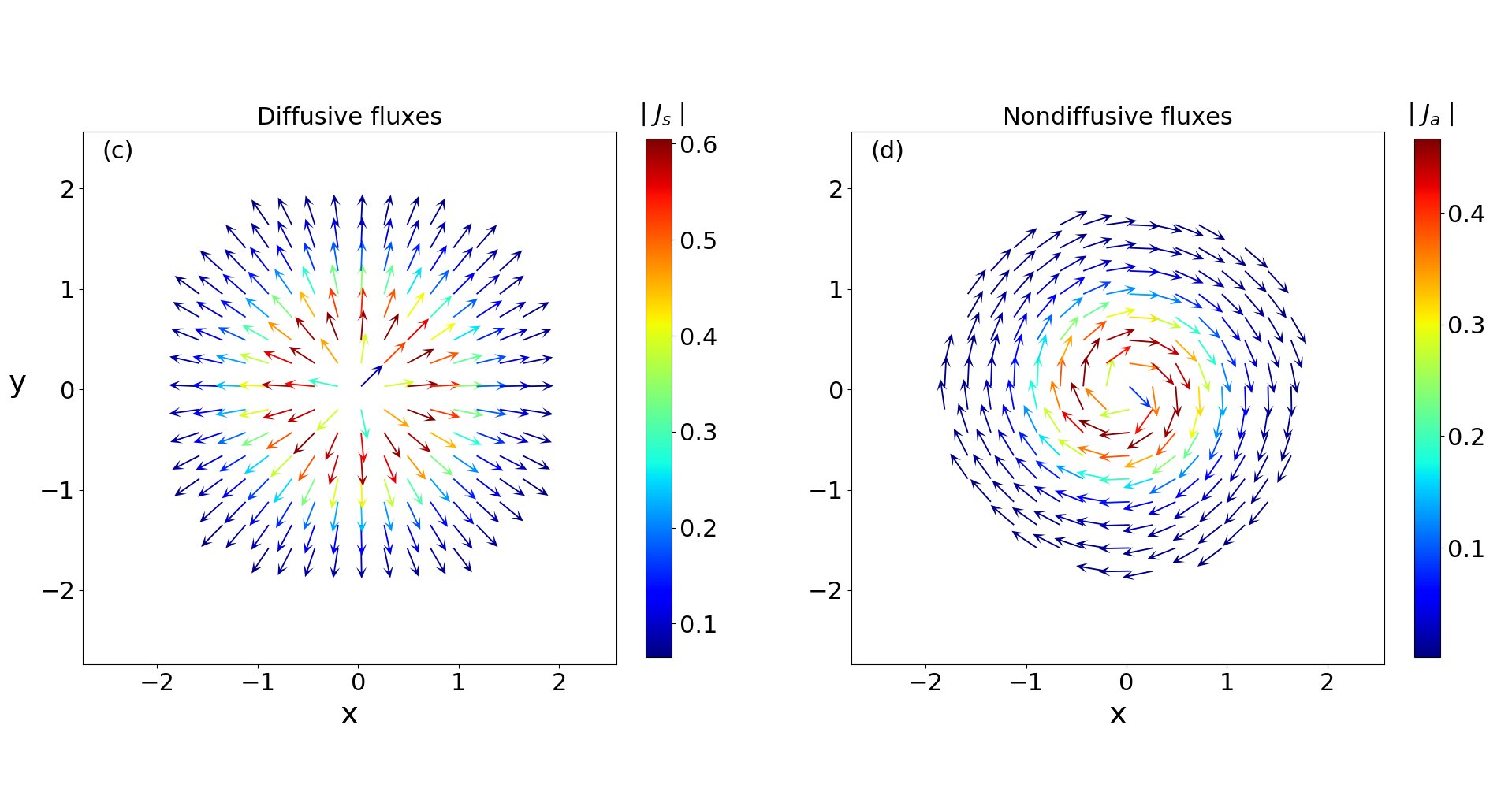}} 
\caption{ Radially symmetric magnetic field. (a) Density distribution in the system at time $t=1.0$ obtained from Eq.~\eqref{FPE} with $\kappa = 4 e^{-r^2}$, where $r$ is the distance from the origin. The system size is $5 \times 5$. (b) Total fluxes in the system (Eq.~\eqref{flux}). The decomposition of the total flux into diffusive and nondiffusive components is shown in (c) and (d), respectively. In this special case, the nondiffusive fluxes do not influence the density evolution. This implies that on retaining only the symmetric part of $\D$ in Eq.~\eqref{FPE}, the same density distribution is obtained as in (a). }
\label{varyingB1}
\end{figure*}
We first consider a system subjected to a constant magnetic field. The particles are initially uniformly distributed in a circle of radius 1 centred at , respectively origin. For a constant magnetic field only the diffusive fluxes contribute to the time evolution of the density probability. This is due to the antisymmetry of $\D_a$ which implies that $\nabla\cdot\J_a(\rr, t) = \D_a\nabla^2P(\rr, t) = 0$, that is the nondiffusive fluxes have a zero-divergence.

Figures~\ref{constantB}(a) and (b) show, respectively, the results for the density and fluxes in the system at time $t = 1.0$. Clearly the flux has radial and rotational components. The radial component corresponds to the diffusive fluxes in the system which exist along the density gradients (see Fig.~\ref{constantB}(c)). The rotational component corresponds to the nondiffusive fluxes which are perpendicular to the density gradients (see Fig.~\ref{constantB}(d)). Since these nondiffusive fluxes are divergence free, they do not affect the density evolution.  
This, however, does not mean that the dynamics of this system are the same as that of a system with only the symmetric tensor. In fact, that there are nondiffusive fluxes present in the system, makes the dynamics distinct from a purely diffusive system.
\begin{figure*}
\centering
\resizebox*{17.5cm}{!}{\includegraphics{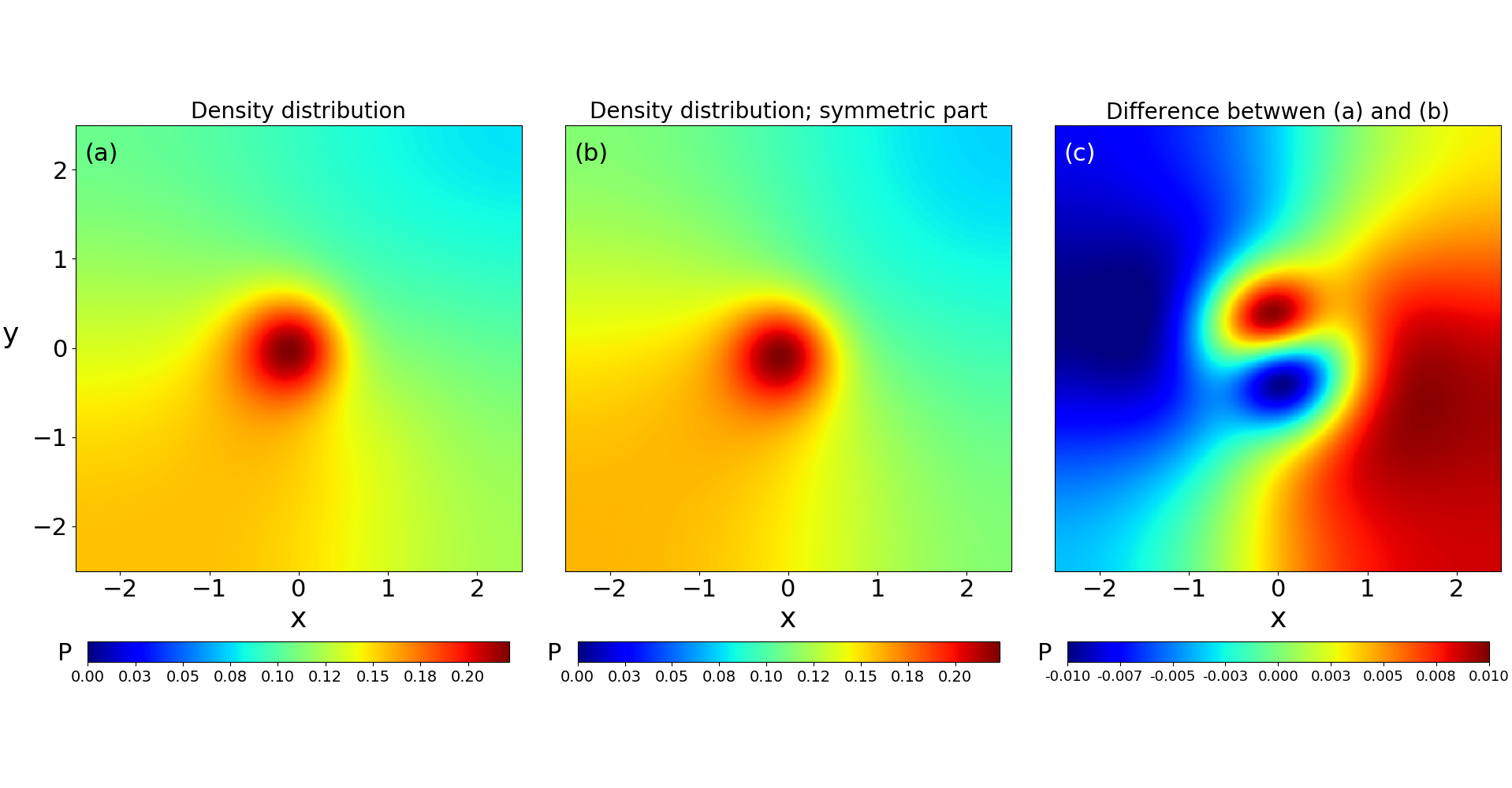}} \\ \vspace*{-12mm}
\resizebox*{17.5cm}{!}{\includegraphics{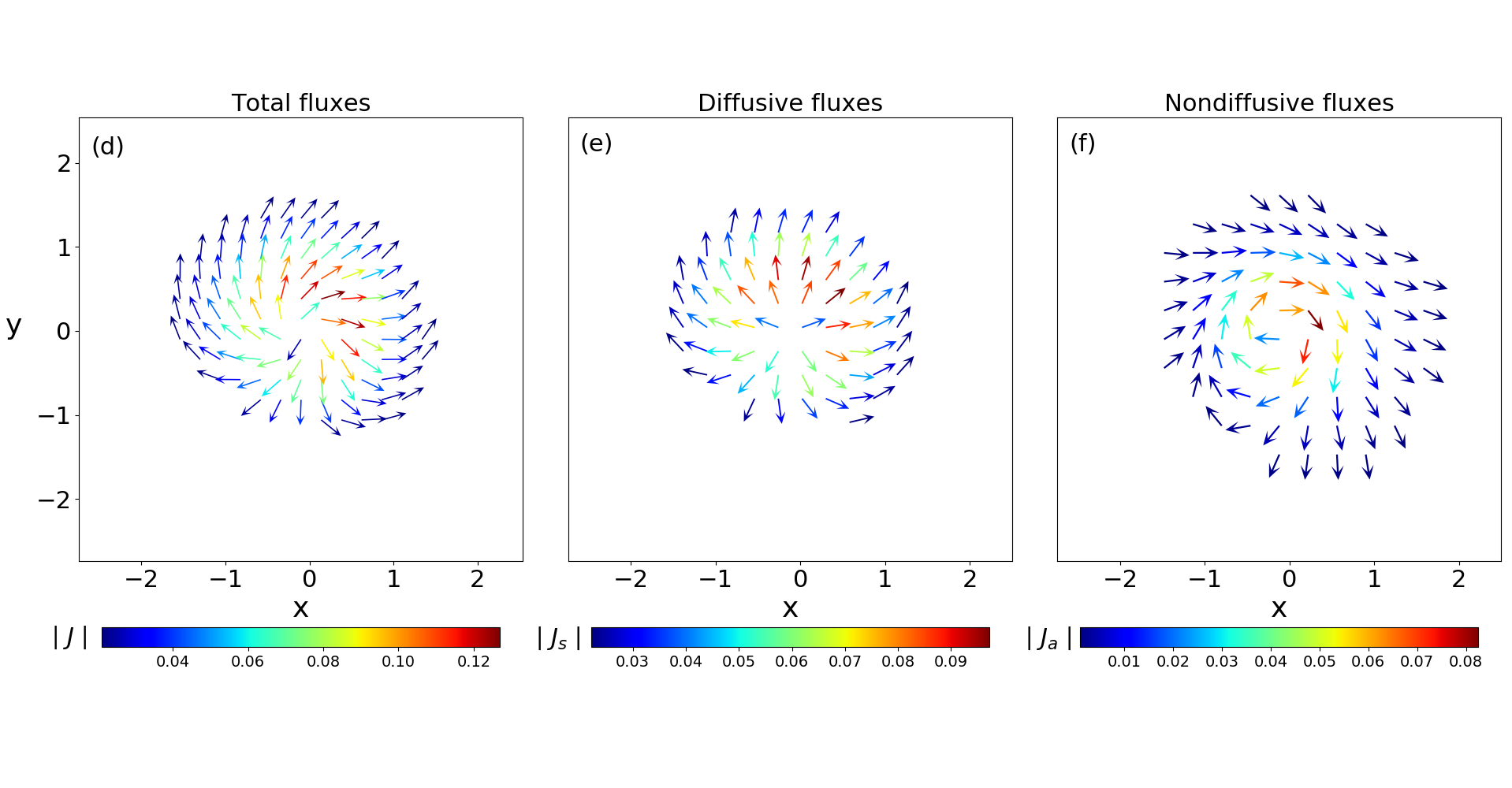}} \\ \vspace*{-10mm}
\caption{ Radially symmetric magnetic field (same as in Fig.~\ref{varyingB1}) and displaced initial condition. Density and fluxes in the system at time $t=4.0$ obtained from Eq.~\eqref{FPE} with $\kappa = 4 e^{-r^2}$ are shown in (a) and (d-f), respectively. The system size is $5\times 5$. (b) The density distribution obtained from dynamics in which only the symmetric part of the tensor is retained. We have slightly displaced the circular initial condition from the origin (in Fig.~\ref{varyingB1}) to (-0.3, -0.3) here. That the two systems have different density distribution is due to the nondiffusive fluxes. This difference between the two density distributions is shown in (c). The decomposition of the total flux into diffusive and nondiffusive components is shown in (e) and (f), respectively. Note that for better visualization, we have chosen larger time here than the previous case.}

\label{varyingB2}
\end{figure*}
For the particular case of constant magnetic field, nondiffusive fluxes can be ignored in studying the density evolution in the system. This has been previously shown by some of the coauthors of this study in Ref.~\cite{vuijk2019anomalous} by integrating the Langevin equation~\eqref{LangevinB} with a small mass. Interestingly, the nondiffusive fluxes are reminiscent of the Corbino effect in conductors~\cite{adams1915hall} (see Fig.~\ref{constantB}(d)): When a disc carrying radial current is subjected to magnetic field in the direction perpendicular to its plane, (additional) circular fluxes are generated. 



\subsection{Radially Symmetric Magnetic Field}
\label{subsectionB}

Figure ~\ref{varyingB1} shows the density and fluxes in the system at time $t = 1.0$ for a radially symmetric, Gaussian shaped magnetic field centred at the origin. The particles are initially uniformly distributed in a circle of radius $1$ centred at the origin.  With this choice, the initial condition and the magnetic field have the same symmetry. This symmetry implies that $P(\rr, t)\equiv P(r, t)$, where $r$ is the distance from the origin. It then follows that $\nabla\cdot\J_a(\rr, t) = \nabla\D_a\cdot\nabla P = 0$ because $\nabla\D_a(\rr)$ is perpendicular to $\nabla P(r, t)$. This means that despite a spatially inhomogeneous magnetic field, the nondiffusive fluxes have no contribution to the evolution of the probability density.
\begin{figure*}
\centering
\vspace*{-1.2cm}
\resizebox*{17.5cm}{!}{\includegraphics{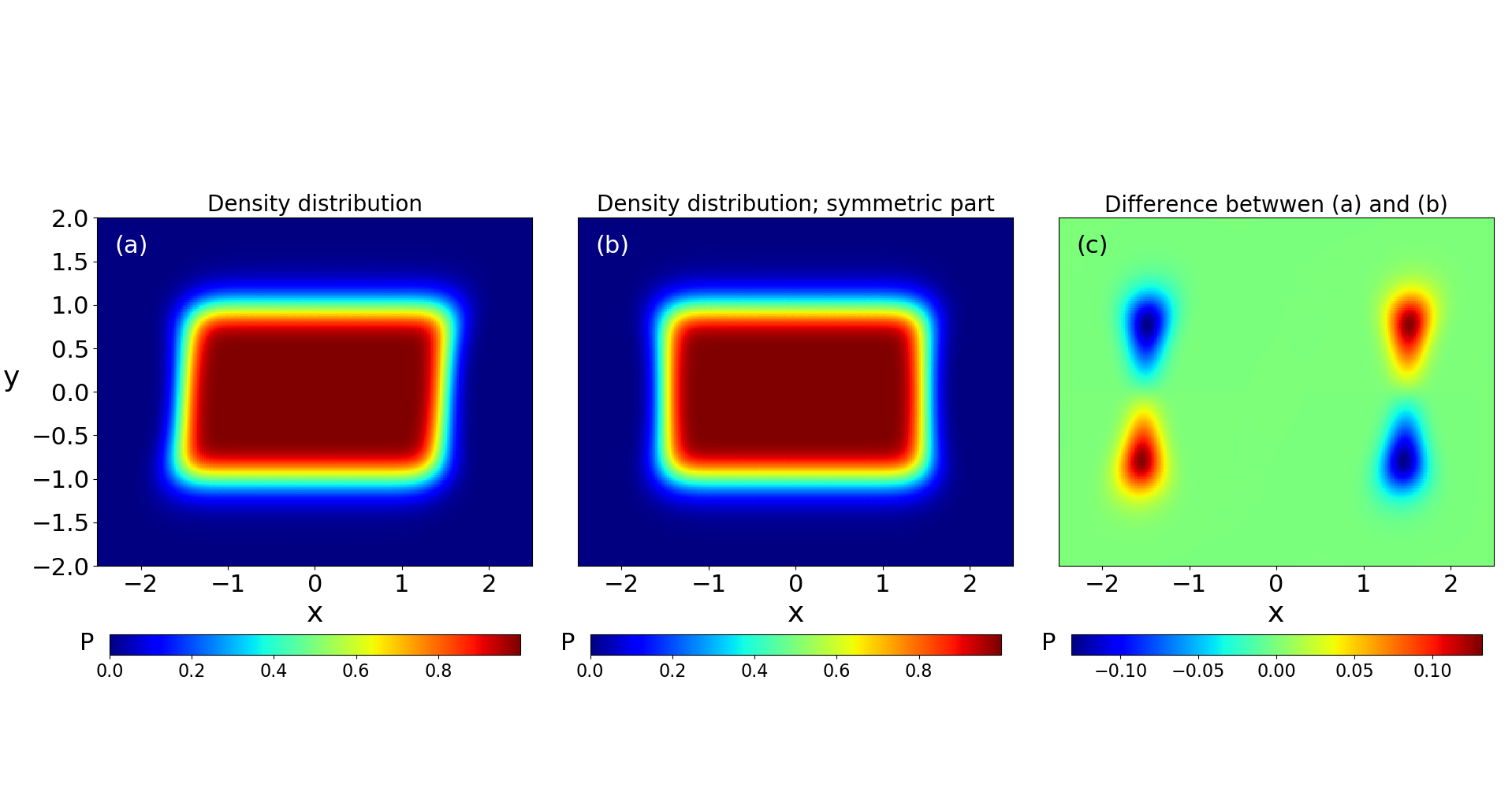}} \\ \vspace*{-12mm}
\resizebox*{17.5cm}{!}{\includegraphics{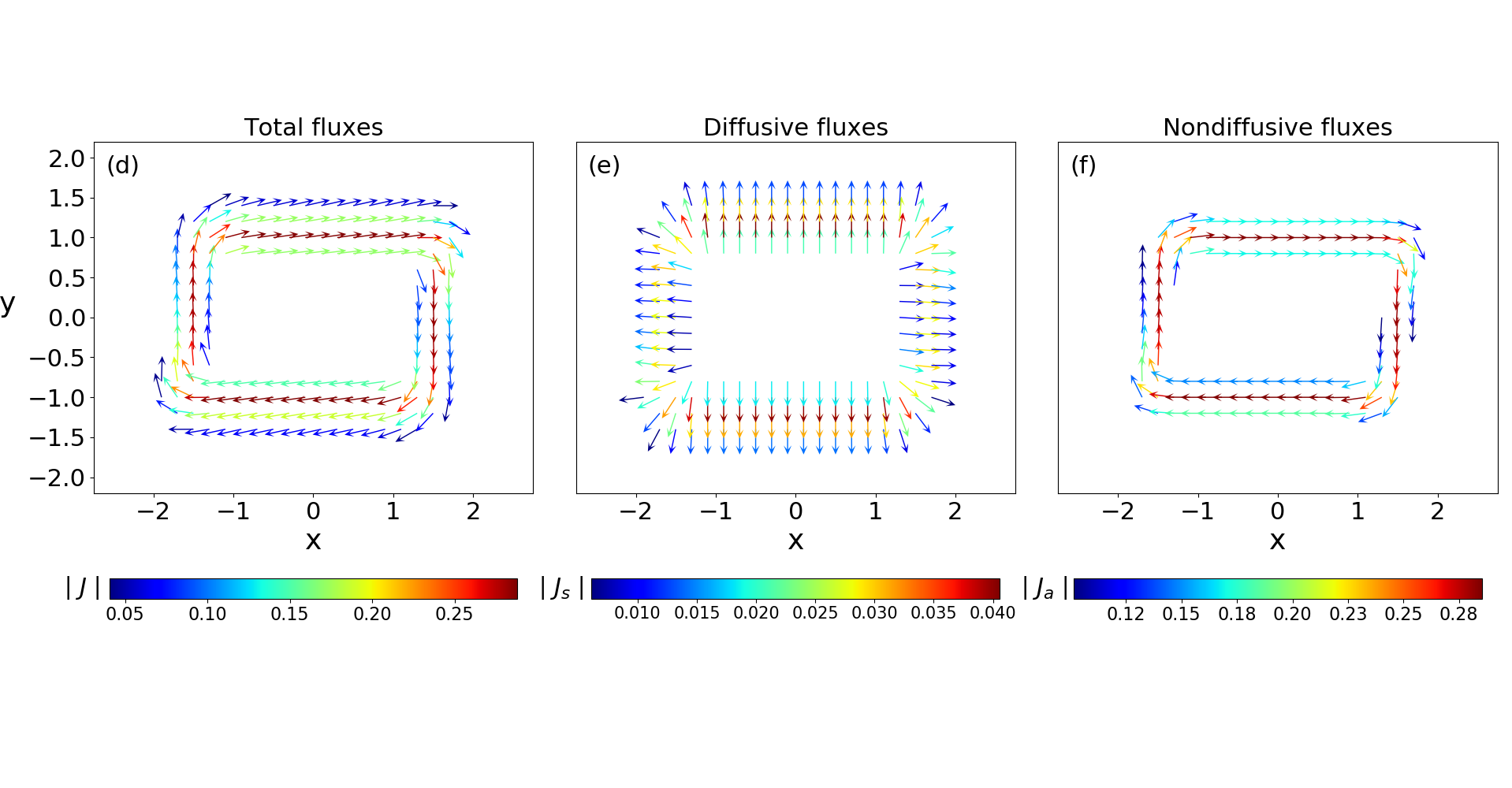}} 
\vspace*{-10mm}
\caption{Magnetic field along $y$ and rectangular initial distribution. Density and fluxes in the system at time $t=1.0$ obtained from Eq.~\eqref{FPE} with $\kappa(y) = - 10\sin(\pi y/4)$ are shown in (a) and (d-f), respectively. The particles are initially uniformly distributed in the region $x\in [-1.5, 1.5]$, $y\in [-1, 1]$ in a system of size $5 \times 4$.  The density distribution becomes distorted due to the nondiffusive fluxes. (b) The density distribution obtained from dynamics in which only the symmetric part of the tensor is retained. Whereas the diffusive fluxes in (e) are perpendicular to the density gradient, nondiffusive fluxes, shown in (f), are parallel to the density gradient. The total fluxes, shown in (d), are predominantly nondiffusive.}
\label{rectangularB}
\end{figure*} 
In this case, one obtains the same density distribution in the system if one considers only the symmetric part of the tensor $\D$. The fluxes, shown in Fig.~\ref{varyingB1}(b), are decomposed into diffusive and nondiffusive parts in Fig.~\ref{varyingB1} (c) and (d), respectively. In this case the time evolution of the density distribution is not influenced by the nondiffusive fluxes.

We now show that on slightly displacing the initial configuration with respect to the magnetic field, the nondiffusive fluxes affect the  time evolution of the density distribution in the system. Whereas the applied magnetic field is the same as in the previous case, the center of the (circular) initial condition is displaced to (-0.3, -0.3). 
Figures~\ref{varyingB2} (a) and (b) show the density of the system at time $t=4.0$ obtained from solving Eq.~\eqref{FPE} with and without the antisymmetric part of $\D$, respectively. The difference between these two is shown in Fig.~\ref{varyingB2} (c) which shows that the nondiffusive fluxes affect the density evolution. The total flux in the system and its decomposition into diffusive and nondiffusive fluxes are shown in Figs.~(d-e).
 




\subsection{Rectangular Initial Distribution}
\label{subsectionC}

Until now we have considered inhomogeneous magnetic field with radial symmetry. We showed that with carefully chosen initial conditions, one can ignore nondiffusive fluxes in the density evolution. We now consider the case in which the magnetic field varies along the $y$ direction $\kappa(y)=-10\sin(\frac{\pi y}{4})$. The particles are initially uniformly distributed in the rectangular region $x\in [-1.5, 1.5]$, $y\in [-1, 1]$ in a system of size $5\times 4$. Figure~\ref{rectangularB} shows the density and fluxes at time $t = 1.0$ obtained from the full dynamics in (a) and (d-f) and from diffusive dynamics in (b), respectively. The total fluxes shown in Fig.~\ref{rectangularB} (d), obtained from full dynamics, however appear to be predominantly nondiffusive in nature. As can be clearly seen in Fig.\ref{rectangularB} (e), diffusive fluxes are parallel to the density gradient. The effect of the nondiffusive fluxes is strikingly evident in the density distribution (see Fig.~\ref{rectangularB} (c)): it becomes distorted under full dynamics in contrast to (b) which remains rectangular.  

Note that the equilibrium density distribution of the system is independent of the applied magnetic field. For any chosen initial configuration and magnetic field, a uniform probability distribution is obtained in the long-time limit. The same holds for a time-dependent magnetic field. In the supplementary material we show a video which shows the fluxes for a time-dependent, spatially constant magnetic field. For a slowly varying magnetic field, the nondiffusive fluxes reverse direction with the polarity of the magnetic field. However, the direction of the diffusive fluxes is constant as the system approaches equilibrium.

\section{Discussion and Conclusion}
\label{section04}
A charged, Brownian particle, subjected to Lorentz force due to an external magnetic field performs anisotropic motion; the rate of diffusion in the plane perpendicular to the magnetic field is reduced whereas the diffusion along the direction of the magnetic field is unchanged. On a statistical level, the anisotropy is encoded in the tensorial coefficient of the Fokker-Planck equation for the probability density of the particle. The tensor, however, is not a diffusion tensor due to the presence of antisymmetric terms\cite{chun2018emergence,vuijk2019anomalous}. This feature which is unique to Lorentz force, gives rise to unusual nondiffusive fluxes in the system. Density gradients in the system not only result in diffusive fluxes along the gradient but also fluxes perpendicular to the gradients. In this paper we studied how these nondiffusive fluxes affect the dynamics of the system. We showed that these fluxes make the dynamics distinct from purely diffusive dynamics. In particular, we showed that for an inhomogeneous magnetic field these nondiffusive fluxes have finite divergence and therefore affect the time evolution of the density. Only in the special cases of a uniform magnetic field or initial condition with the same symmetry as the magnetic field, can these fluxes be ignored in the density evolution.

There are several interesting directions in which this work can be extended. The Fokker-Planck approach can be generalised to a system of interacting particles subjected to an inhomogeneous magnetic field. It is then straightforward to derive a coarse-grained equation for the one-body density and its time evolution within the framework of dynamical density functional theory~\cite{archer2004dynamical, vuijk2019effect}. It will be interesting to study how nondiffusive fluxes affect the phase transition dynamics of a fluid system. It has recently been shown that Lorentz force can induce unusual nonequilibrium steady state in a system of active Brownian particles~\cite{vuijk2019lorenz}. It would be interesting to study the dynamics leading to the nonequilibrium steady state in such a system. 

\providecommand{\noopsort}[1]{}\providecommand{\singleletter}[1]{#1}%

\end{document}